\documentclass[10,conference]{IEEEtran}
\usepackage{amsmath}
\usepackage{amsfonts}
\usepackage{amssymb}
\usepackage{amscd}
\usepackage[dvips]{graphicx}
\usepackage{epsfig}
\usepackage{subfigure}
\usepackage{newlfont}
\usepackage{cite}
\usepackage{subfigure}
\usepackage{enumitem}
\IEEEoverridecommandlockouts
\newcommand{\setxysizeo}{\epsfxsize=3in}
\newcommand{\pullUp}{\vskip -0.2cm}

\newcommand{\lsqb}{\left[}
\newcommand{\rsqb}{\right]}
\newcommand{\lcb}{\left\{}
\newcommand{\rcb}{\right\}}
\begin{document}

\title{Secrecy in the $2$-User Symmetric Deterministic Interference Channel with Transmitter Cooperation} 

\author{\IEEEauthorblockN{Parthajit Mohapatra and Chandra R. Murthy}\vspace{-1cm}
\thanks{The authors are with the Dept. of ECE, Indian Institute of Science, Bangalore 560 012, India (e-mails:\{partha, cmurthy\}@ece.iisc.ernet.in).}}


\maketitle

\begin{abstract}
This work presents novel achievable schemes for the $2$-user symmetric linear deterministic interference channel with limited-rate transmitter cooperation and perfect secrecy constraints at the receivers. The proposed achievable scheme consists of a combination of interference cancelation, relaying of the other user's data bits, time sharing, and transmission of random bits, depending on the rate of the cooperative link and the relative strengths of the signal and the interference. The results show, for example, that the proposed scheme achieves the same rate as the capacity without the secrecy constraints, in the initial part of the weak interference regime. Also, sharing random bits through the cooperative link can achieve a higher secrecy rate compared to sharing data bits, in the very high interference regime. The results highlight the importance of limited transmitter cooperation in facilitating secure communications over $2$-user interference channels. 
\end{abstract}
\section{Introduction}
In multiuser wireless communications, users experience interference due to the broadcast and superposition nature of the medium. Interference not only limits the performance of the system, but also allows users to eavesdrop on the other users' messages. For example, in a cellular network, when users have subscribed to different contents, it is important for the service provider to support high throughput, as well as secure its transmissions, in order to maximize its own revenue.  In these scenarios, the transmitters (e.g., base stations) are not completely isolated from each other, and cooperation among them is possible. Such cooperation can potentially provide significant gains in the achievable throughput in the presence of interference, while simultaneously guaranteeing security.  In this work, we investigate the effectiveness of limited transmitter cooperation in a $2$-user symmetric linear deterministic interference channel (SLDIC) on interference management and secrecy. 

Information theoretic secrecy in the interference channel (IC) with $K\geq 2$ users and different eavesdropper settings have been analyzed in \cite{hgamal1,liu1}. In \cite{hgamal1}, the secrecy in $2$-user IC is considered in the presence of an eavesdropper. In \cite{liu1}, the broadcast and the IC with independent confidential messages are considered, and optimality is established in some special cases. The frequency/time selective $K$-user Gaussian IC (GIC) with secrecy constraints is considered in \cite{hgamal2}. The effect of cooperation on secrecy has been explored in \cite{ekrem2,ekrem1,awan1}. 
A linear deterministic model  for relay network was introduced in \cite{avesti1}, which led to insights on the achievable schemes in Gaussian relay networks. The deterministic model has subsequently been used for studying the achievable rates with the secrecy constraints in \cite{yates1, perron1, shamai1}. However, the effect of limited transmitter-side cooperation on secrecy in an IC has not been explored in literature, and is the focus of this work.

In this paper, novel transmission schemes for the $2$-user SLDIC with limited transmitter cooperation and secrecy constraints at the receivers are proposed, and their achievable secrecy rates are derived. The transmission scheme depends on the capacity of the cooperative link (denoted by $C$) and value of $\alpha \triangleq \frac{n}{m}$, where $m \triangleq (\lfloor \log \text{\textsf{SNR}}\rfloor)^{+}$ and $n \triangleq  (\lfloor \log \text{\textsf{INR}}\rfloor)^{+}$. The key features of the proposed schemes are:
\begin{enumerate}
\item In the weak interference regime\footnote{Note that the definition of the weak interference regime here is different from the more typical $(0 < \alpha \leq \frac{1}{2})$\cite{etkin1}. It will turn out that $(0 < \alpha \leq \frac{2}{3})$ is more appropriate for the discussion in this paper.}
$(0 < \alpha \leq \frac{2}{3})$, the scheme involves precoding of a user's own data bits with the  bits received through cooperation, to simultaneously cancel the interference and ensure secrecy. 
\item In the moderate interference regime $(\frac{2}{3} < \alpha < 1)$, the scheme uses interference cancelation, random bit transmission, or both. The novel idea behind the  random bit transmission scheme is explained in Sec.~\ref{sec:modint}. 
\item In the high interference regime $(1 < \alpha < 2)$, the  scheme involves relaying of the other user's data bits obtained at the transmitters through the cooperative links, in addition to the techniques used for~$(\frac{2}{3} < \alpha < 1)$. 
\item In the very high interference regime $(\alpha \geq 2)$, 
the scheme uses time sharing, along with 
the techniques used for $(1 < \alpha < 2)$. 
Unlike the other interference regimes, when $\alpha \geq 2$ and for small values of $C$, sharing random bits along with the data bits is strictly better that sharing only data bits, in terms of the achievable secrecy rate. 
\end{enumerate}
To the best of the authors' knowledge, the transmission schemes proposed in this work have not hitherto been studied in the literature. Further, the secrecy rate achievable by the proposed schemes is derived. It is shown that it is possible to achieve a nonzero secrecy rate in almost all cases, with limited transmitter cooperation. In some cases, the achievable secrecy rate equals the capacity of the same system without the secrecy constraints. Thus, the proposed schemes allow one to get secure communications for free, in these cases. 

Due to lack of space, the proposed transmission schemes and the corresponding achievable rates are stated only for some specific interference regimes. The complete details will be provided in the extended version of this work~\cite{partha3}. 

Notation: Lower case or upper case letters are used to represent scalars. Small boldface letters represent vectors, whereas capital boldface letters represent matrices. 
\section{System Model}\label{sec:sysmod}
Consider a two user Gaussian symmetric IC (GSIC) with cooperating transmitters. The  signals at the receivers are modeled as
\begin{align}
 y_1 = h_{d}x_{1} + h_{c}x_2 + z_{1}; \quad
 y_2 = h_{d}x_{2} + h_{c}x_1 + z_{2}, \label{sysmodel1}
\end{align}
where $z_{j}\:(j=1,2)$ is complex Gaussian, distributed as $z_{j} \sim \mathcal{CN}(0,1)$. Here, $h_d$ and $h_c$ are the channel gains of the direct and cross links, respectively. The transmitters cooperate through a noiseless link of finite rate $(C_{G})$. The equivalent deterministic model of (\ref{sysmodel1}) at high SNR is as follows \cite{wang1}:
\begin{align}
\mathbf{y}_{1} = \mathbf{D}^{q-m}\mathbf{x}_{1} \oplus \mathbf{D}^{q-n}\mathbf{x}_{2}; \  \mathbf{y}_{2} = \mathbf{D}^{q-m}\mathbf{x}_{2} \oplus \mathbf{D}^{q-n}\mathbf{x}_{1}, \label{sysmodel2}
\end{align}
where $\mathbf{x}_{i}$ and $\mathbf{y}_{i}$ are binary vectors of length $q \triangleq \max\{m,n\}$, $\mathbf{D}$ is a $q \times q$ downshift matrix with elements $d_{j,k}=1$ if $2 \leq j=k+1\leq q$ and $d_{j,k}=0$ otherwise, and $\oplus $ stands for  modulo-$2$ addition (\textsf{XOR} operation). The parameters $m$ and $n$ are related to the GSIC as $ m = (\lfloor \log|h_d|^2\rfloor)^{+},\: n =  (\lfloor \log|h_c|^2\rfloor)^{+},$  while the capacity of the cooperative link is $C = \lfloor C_{G} \rfloor$ \cite{wang1}. 
The quantity $\alpha \triangleq \frac{n}{m}$ captures the amount of coupling between the signal and the interference, and is central to characterizing the achievable rates of the LDIC.

The convention followed for the LDIC is the same as that presented in \cite{wang1}. The bits $a_{i}, b_i \in \{0,1\}$ denote the  information bits of transmitters $1$ and $2$, respectively, sent on the $i^{\text{th}}$ level, \emph{with the levels numbered starting from the bottom-most entry}. The bits transmitted on the different levels of the LDIC are chosen to be equiprobable Bernoulli distributed, denoted by $\operatorname{Bern} \left({\frac{1}{2}}\right)$. The bits $d_{i}, e_i \in \{0,1\}$ denote the random bits generated by transmitters $1$ and $2$, respectively, and sent on the $i^{\text{th}}$ level. The random bits are independent of the data bits, and are generated from the $\operatorname{Bern} \left({\frac{1}{2}}\right)$  distribution.

The transmitter $i$ has a message $W_{i}$, which should be decodable at the intended receiver $i$, but needs to be kept perfectly secret from the other, unintended receiver $j$, $j \neq i$. The encoded message  is a function of its own data bits, the bits received through the cooperative link, and possibly some random data bits. 
The encoding at the transmitter should satisfy the causality constraint, i.e., it cannot depend on future cooperative bits. 
The decoding is based on solving the linear equation in \eqref{sysmodel2} at each receiver. Also, it is assumed that the transmitters trust each other completely and that they do not deviate from the agreed scheme. For secrecy, it is required to satisfy $I(W_{i}, \mathbf{y}_{j}) = 0, i,j \in \{1,2\} \text{ and } i \neq j$~\cite{shannon1}.

\section{Achievable Schemes}

\subsection{Weak interference regime $(0 \leq \alpha \leq \frac{2}{3})$:}
When $0 \leq \alpha \leq \frac{2}{3}$, bits transmitted on the lower $m-n$ levels $[1:m-n]$ do not cause interference at the unintended receiver, and data bits transmitted on these levels will remain secure. But, the bits transmitted on the top $n$ levels $[m-n+1:m]$ will cause interference at the unintended receiver at the bottom $n$ levels $[1:n]$. Without transmitter cooperation, if all the bottom $m-n$ levels are used for transmission, it is easy to see that transmitting on the remaining  $n$ levels either reduces the rate or violates secrecy. Hence, it is possible to transmit $m-n$ bits securely, when $C=0$. 

\begin{figure}
\centering
\mbox{\subfigure[][]{\includegraphics[width=1.65in]{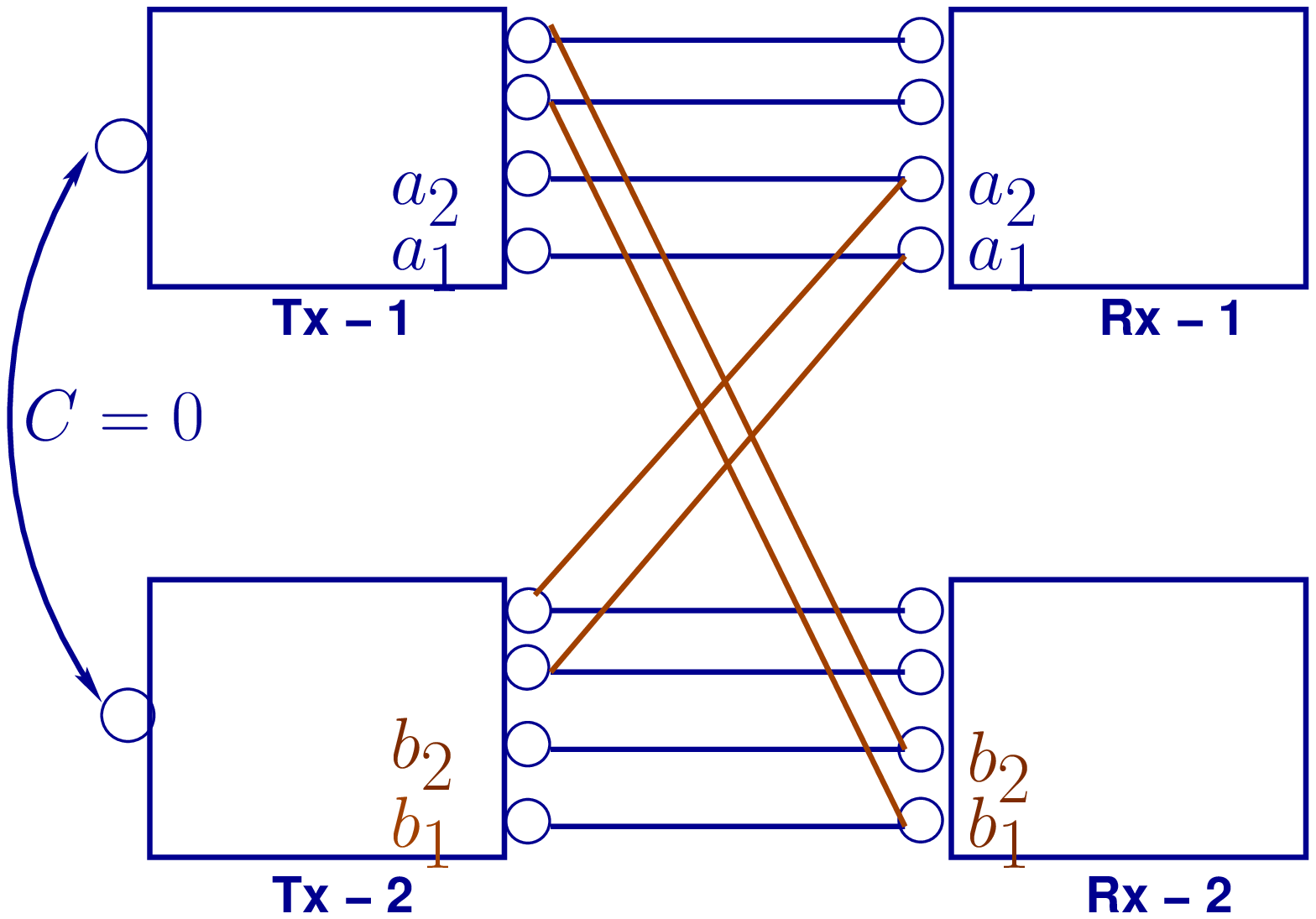}\label{fig:weak1}} \quad
\subfigure[][]{\includegraphics[width=1.65in]{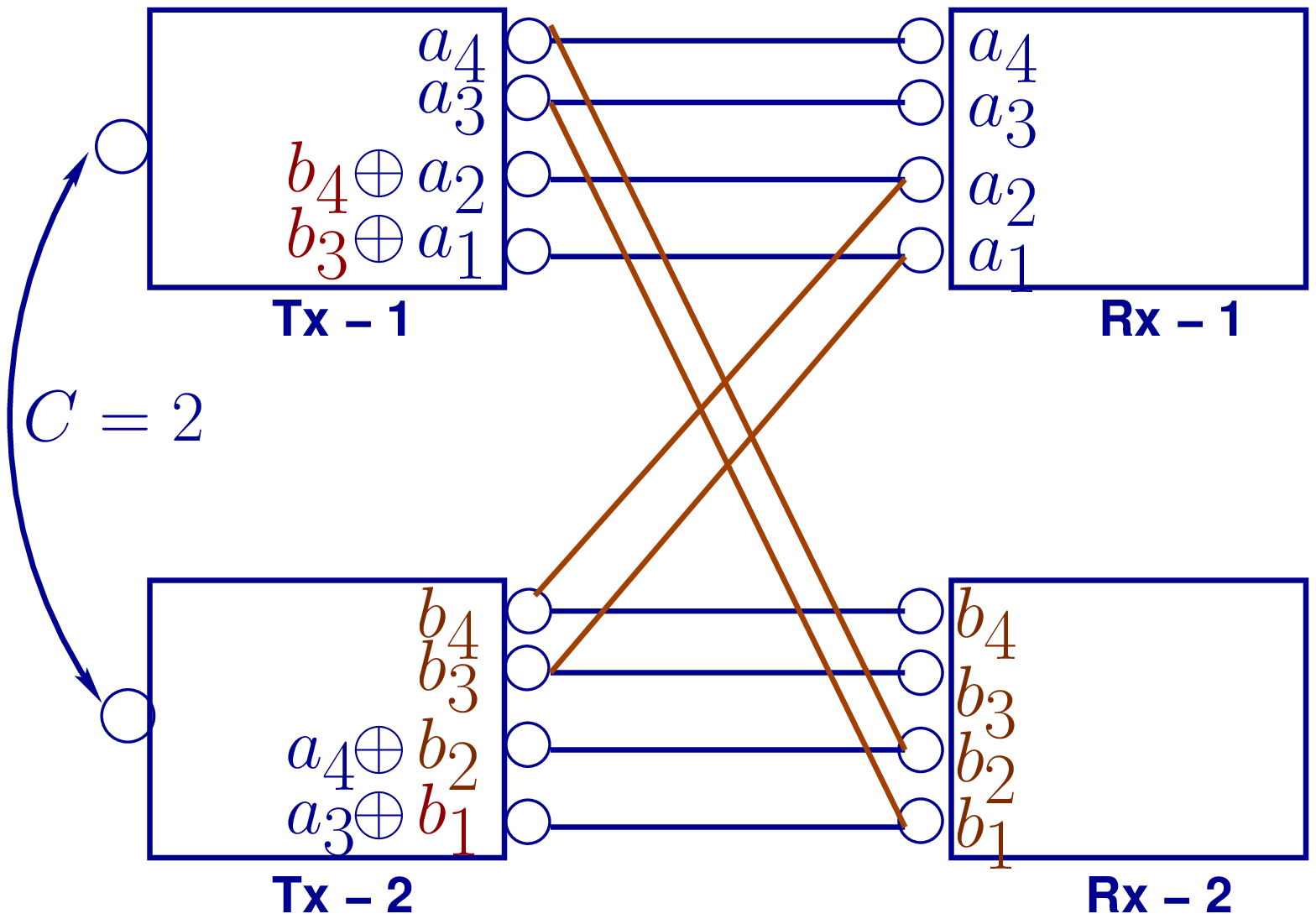}\label{fig:weak3}}}
\caption[]{LDIC with $m=4$ and $n=2$: \subref{fig:weak1} $C=0$, $R_S=2$ and \subref{fig:weak3} $C=2$, $R_S=4$.}\label{fig:weak4}
\pullUp
\end{figure}

With transmitter cooperation ($C>0$), it is possible to transmit on the top levels by appropriately xoring the data bits with the cooperative bits in the lower levels prior to transmission, as follows. The transmitters exchange $\min\{n,C\}$ bits, which are the bits they intend to transmit on the levels $[m-n + 1:m-n+\min\{n,C\}]$, through the cooperative link. These cooperative bits are precoded with the data bits at the levels $[1:\min\{n,C\}]$ to cancel interference caused by the data bits sent by the other transmitter. This scheme is illustrated for $C=0$ and $2$ in Fig.~\ref{fig:weak4}. Mathematically, when $C \leq n$,
the message of transmitter $1$ is encoded as follows:
\begin{align}
\mathbf{x}_{1} = \lsqb\begin{array}{l} 
                  \mathbf{0}_{(m-(r+C))^{+} \times 1} \\
                  \mathbf{a}_{(r+C) \times 1} \end{array}\rsqb \oplus
                  \lsqb\begin{array}{l}
                  \mathbf{0}_{(m-C) \times 1} \\
                  \mathbf{b}_{C \times 1}^{c} \end{array}\rsqb, \label{weakach1}
\end{align}
where $\mathbf{a} \triangleq [a_{r+C},a_{r+C-1},\ldots,a_1]^{T}$ are the own data bits, $\mathbf{b}^{c} \triangleq [b_{r+C}, b_{r+C-1},  \ldots, b_{r+1}]^{T}$ are the cooperative data bits, and $r \triangleq m-n$. The message of transmitter $2$ is encoded in an analogous fashion. When $C=n$, it can be shown that the proposed scheme achieves $\max\{m,n\}$, the maximum rate possible in the LDIC. When $C>n$, $C-n$ bits can be discarded and $n$ cooperative bits can be used for encoding. Hence, in the sequel, it will not be explicitly mentioned that $C \leq n$. The proposed encoding scheme achieves the following symmetric secrecy rate:
\begin{align}
R_{S} = m-n + \min\{n,C\}.  \label{weakach2}
\end{align}

\subsection{Moderate interference regime $(\frac{2}{3} < \alpha < 1)$:}\label{sec:modint}
\begin{figure}
\centering
\mbox{\subfigure[][]{\includegraphics[width=1.51in]{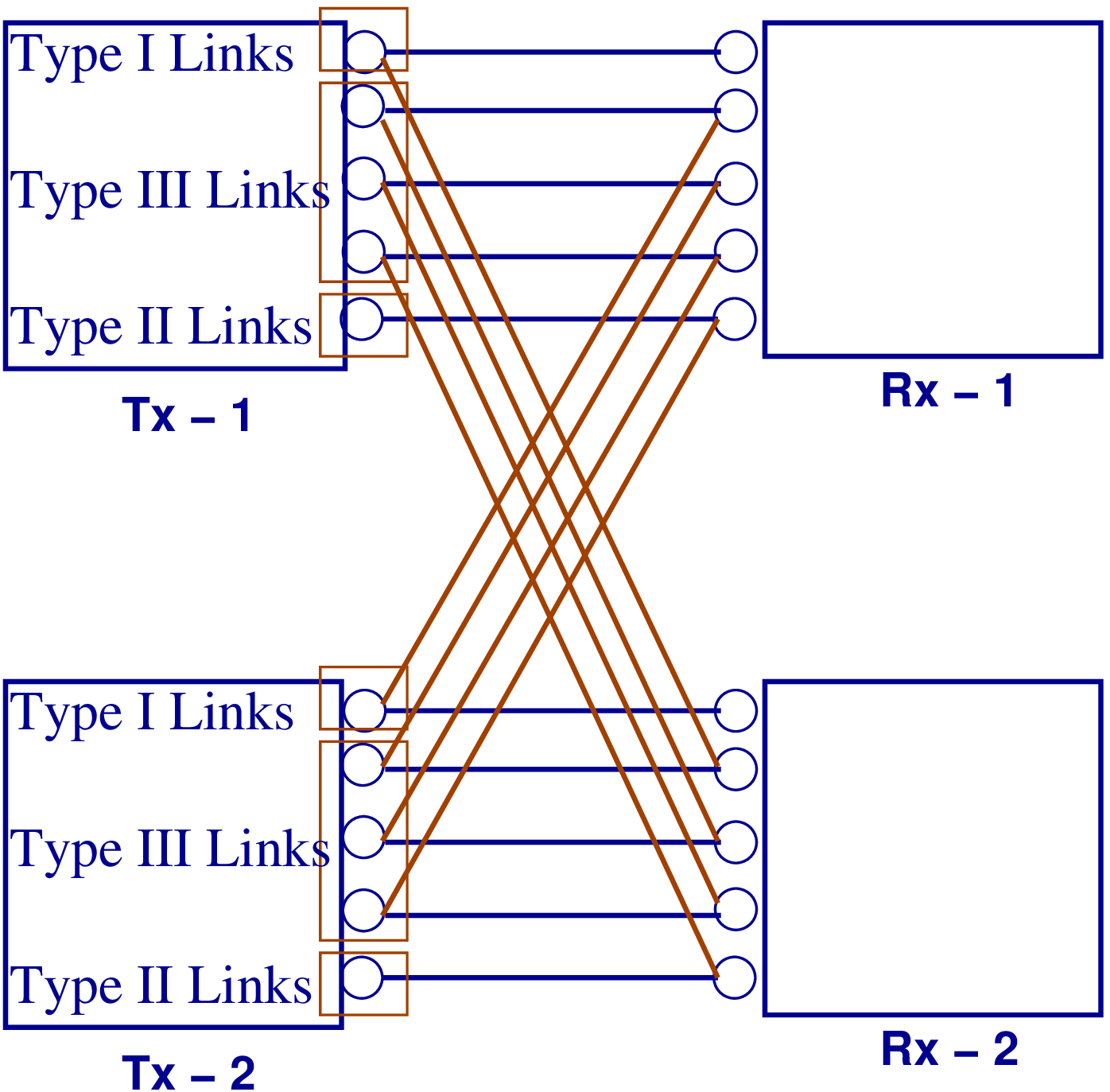}\label{fig:weakmod2}} \quad
\subfigure[][]{\includegraphics[width=1.65in]{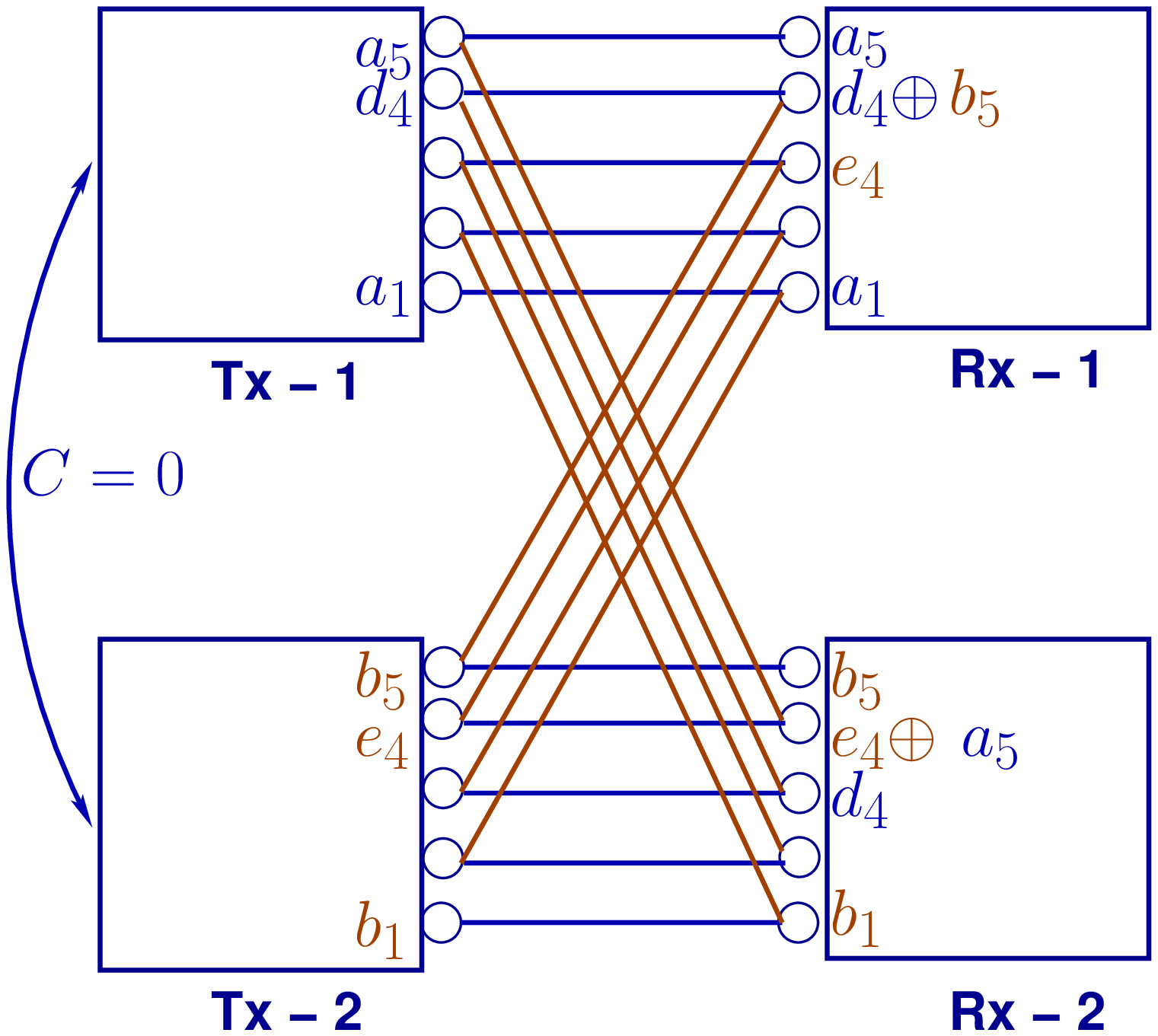}\label{fig:weakmod3}}} 
\mbox{\subfigure[][]{\includegraphics[width=1.65in]{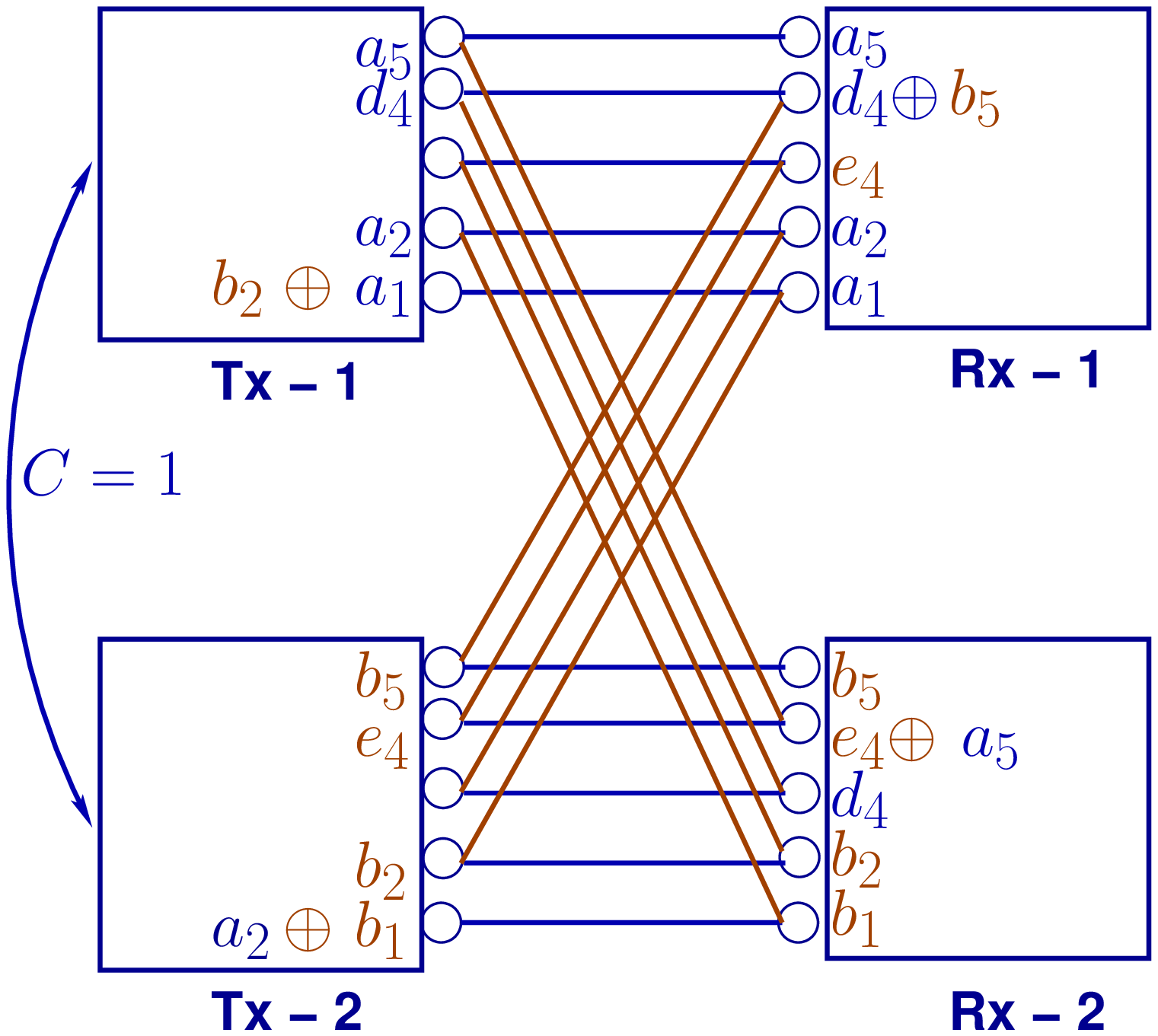}\label{fig:weakmod4}} \quad
\subfigure[][]{\includegraphics[width=1.65in]{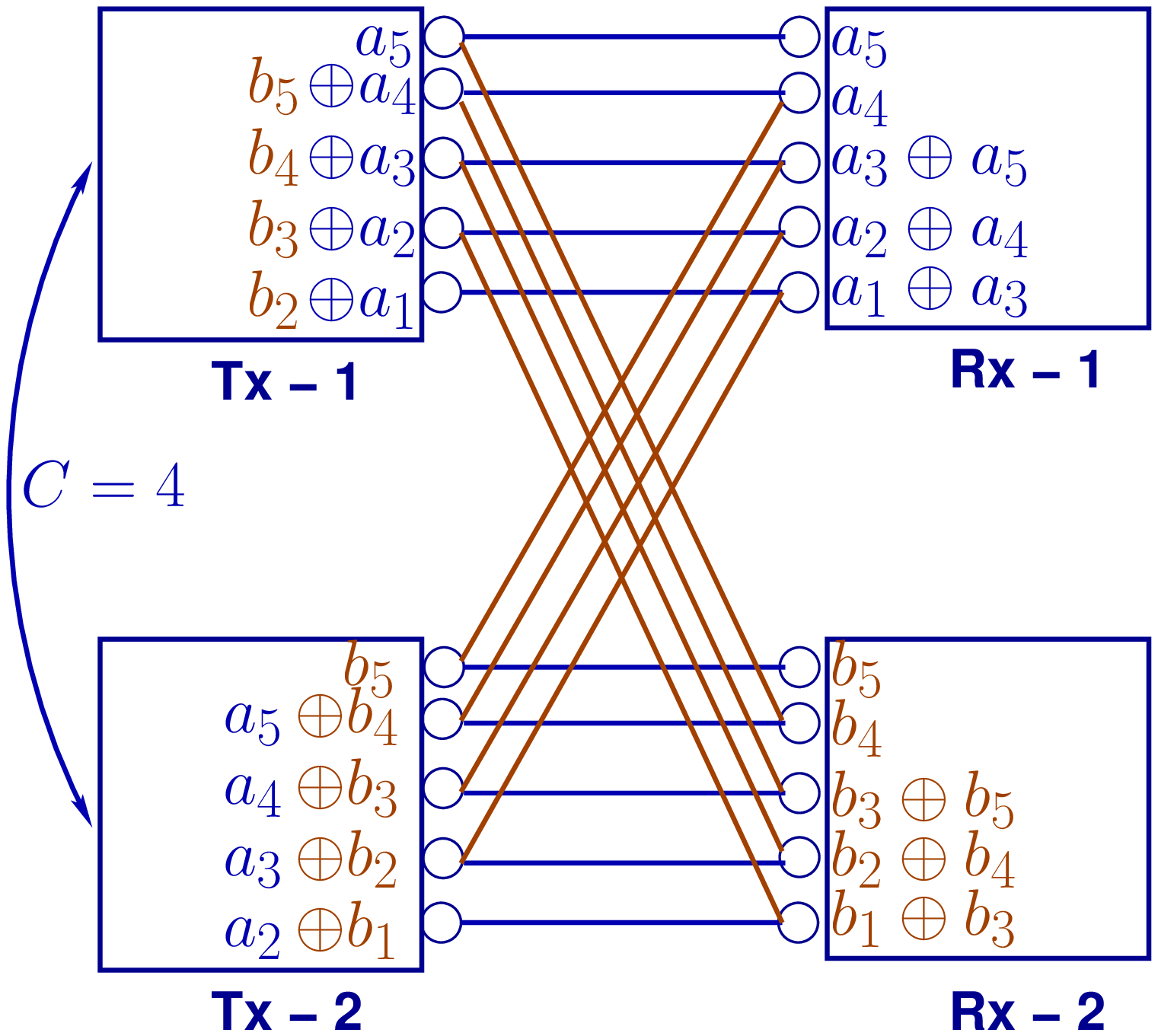}\label{fig:weakmod5}}} \\
\caption[]{LDIC with $m=5$ and $n=4$: \subref{fig:weakmod2} Different types of links, \subref{fig:weakmod3} $C=0$, $R_S=2$ \subref{fig:weakmod4} $C=1$, $R_S=3$, and \subref{fig:weakmod5} $C=4$, $R_S=5$.}\label{fig:weakmod6}
\pullUp
\end{figure}

First, note that the links in the LDIC can be classified into three categories: Type I, Type II, and Type III, as shown in Fig. \ref{fig:weakmod2}. The classification is based on whether the data bits are received cleanly or with interference at the intended receiver, and whether or not they cause interference at the unintended receiver. The number $(r_1)$ of Type I and $(r_2)$ of Type II links are $r_1 = r_2 = \frac{m-l}{2} = m-n$, where $l$ is the number of Type III links, which is given by $l = 2n-m$. 
 
First consider $C=0$. As the bits transmitted on the Type II links $[1:m-n]$ are not received at the unintended receiver, it is possible to transmit at least $r_2$ bits securely. Data bits transmitted on the Type III/I links (levels $[m-n+1:n]/[n+1:m]$) will cause interference at the unintended receiver, and it is not possible to ensure secrecy with uncoded data transmission on these levels. As the Type II links are already used up for data transmission, the remaining $g \triangleq \lcb n-(r_2 + C)\rcb^{+}$ levels\footnote{Although $C=0$ here, the expressions are written including $C$ for use in the sequel.} can be used for transmission with the help of random bits sent by each transmitter. Transmitter $i$ sends the random bits in such a way that they superimpose with the data bits sent by the other transmitter, at receiver $i$. Note that, the receiver does not require the knowledge of these random bits in order to decode its own message. 

Now, it is required to determine the number of levels of Type I/III links that can be used for data transmission. Notice that bits transmitted on any level get shifted down by $m-n$ levels at the unintended receiver. In the proposed scheme,  transmission occurs in blocks of size $3(m-n)$ levels, with each block consisting of a sequence of data bits, random bits and zero-bits of size $m-n$ each, sent on consecutive levels. Such a  scheme ensures that the intended data bits are received cleanly at the desired receiver, and, data received at the unintended receiver remains secure. The total number of blocks of size $3(m-n)$ that can be sent is $B \triangleq \left\lfloor\frac{g}{3(m-n)} \right\rfloor$.  
The remaining $t\triangleq g\% \{3(m-n)\}$ levels may or may not be usable for data transmission depending on the number of levels remaining for random bit transmission. The quantity $q = \min \lcb (t-r_2)^{+}, r_2\rcb$ is the number of data bits that can be securely sent on the remaining $t$ levels. With a little bookkeeping, it can be shown that, for $C=0$, the proposed scheme achieves the following symmetric secrecy rate:
\begin{align}
R_S = m-n + B(m-n) + q. \label{eq:modach1}
\end{align}

When $C>0$, the achievable scheme uses interference cancelation in addition to random bit transmission. Bits transmitted on the levels $[r_2 + 1:r_2 + C]$ at transmitter $i$ will interfere with the levels $[1:C]$ of receiver $j$. The interference can be eliminated by precoding the data bits at levels $[1:C]$ at transmitter $j$  with the data bits of transmitter $i$, received through cooperation. 
The remaining $g$ levels can be used for data transmission with random bit transmission, exactly as in the $C=0$ case. The achievable scheme is illustrated for different values of $C$ in Figs. \ref{fig:weakmod3}, \ref{fig:weakmod4} and \ref{fig:weakmod5}.
Mathematically, the message $\mathbf{x}_1$ of transmitter $1$  is encoded as follows:\\
\textit{Case 1 $(q=0)$:}
\begin{align}
\mathbf{x}_{1} = \lsqb\begin{array}{l}
                  \mathbf{0}_{(m - (r_2 + C))^{+} \times 1} \\
                  \mathbf{a}_{(r_2 + C) \times 1} \end{array}\rsqb \oplus
                  \lsqb\begin{array}{l}
                  \mathbf{0}_{(m-C) \times 1} \\
                  \mathbf{b}_{C \times 1}^{c} \end{array}\rsqb \oplus 
		  \lsqb\begin{array}{l}
                  \mathbf{a}_{p \times 1}^{u} \\
                  \mathbf{0}_{p' \times 1} \end{array}\rsqb,
		  \label{eq:weakmod1}
\end{align}
where $\mathbf{a} \triangleq \lsqb a_{r_2 + C}, a_{r_2 + C-1},\ldots,a_1\rsqb^{T}$, $\mathbf{b}^{c} \triangleq \lsqb b_{r_2 + C},  \right. \\ \left. b_{r_2 + C-1}, \ldots, b_{r_2 + 1}\rsqb^{T}$, $\mathbf{a}^{u} \triangleq \lsqb \mathbf{u}_1, \mathbf{d}_2, \mathbf{z}_3, \mathbf{u}_4, \mathbf{d}_5, \mathbf{z}_6, \ldots, \right. \\ \left. \mathbf{u}_{3B-2}, \mathbf{d}_{3B-1}, \mathbf{z}_{3B},\rsqb^{T}$, $\mathbf{u}_l \triangleq \lsqb a_{m-(l-1)r_2}, a_{m-(l-1)r_2-1},  \right. \\ \left. \ldots, a_{m-lr_2+1}\rsqb$, $\mathbf{d}_l \triangleq \lsqb d_{m-(l-1)r_2}, d_{m-(l-1)r_2-1}, \ldots, \right. \\ \left. d_{m-lr_2+1} \rsqb$, $\mathbf{z}_l$ is a zero vector of size $1 \times r_2$, $p \triangleq 3B(m-n)$ and $p' \triangleq m-p$. The encoding at transmitter $2$ is similar.\\
\textit{Case 2 $(q \neq 0)$:} The number of data bits that can be sent on the remaining  $t$ levels is $q= \min\{(t-r_2)^{+},r_2\}$. In this case, the message of transmitter $1$ is encoded as follows.
\begin{align}
\mathbf{x}_{1}^{\text{mod}} = \mathbf{x}_{1} \oplus 
	      \lsqb\begin{array}{l}
	           \mathbf{0}_{p \times 1} \\
		    \mathbf{a'}_{t \times 1} \\
                   \mathbf{0}_{(m-(p+t)) \times 1}
	           \end{array} \rsqb \label{eq:weakmod2}
\end{align}
where $\mathbf{x}_1$ is as defined in (\ref{eq:weakmod1}), $\mathbf{a'} \triangleq \lsqb \mathbf{u}_{11}, \mathbf{u}_{12}, \mathbf{d}_{11}, \mathbf{d}_{12}, \mathbf{z}\rsqb^{T}$, $\mathbf{u}_{11} = \lsqb a_{m-p}, a_{m-p-1}, \ldots, a_{m-p-q+1}\rsqb$, $\mathbf{d}_{11} \triangleq \lsqb d_{m-w},\right. \\ \left. d_{m-w-1}, \ldots, d_{m-w-q+1}\rsqb$. Also, $\mathbf{u}_{12}$, $\mathbf{d}_{12}$ and $\mathbf{z}$ are zero vectors of size $1 \times v$,  $1 \times f$,  and $1 \times v'$,  respectively. Here, $v \triangleq  (r_2-q)$, $f \triangleq (t-(r_2+q))^{+}$, $w \triangleq p+r_2$ and $v' \triangleq (t-2r_2)^{+}$. 

The proposed encoding scheme achieves the following symmetric secrecy rate:
\begin{align}
R_S  = m-n + B(m-n) + q + \min\{n,C\}. \label{eq:weakmod3}
\end{align}

\subsection{Interference is as strong as the signal $(\alpha=1)$:} In this case, from (\ref{sysmodel2}), $\mathbf{y}_1=\mathbf{y}_2 = \mathbf{x}_{1} \oplus \mathbf{x}_{2}$. Hence, it may not be possible to achieve a nonzero  secrecy rate, as both the receivers see the same signal.

\subsection{Very high interference regime $(\alpha \geq 2)$:}
Due to lack of space, the achievable scheme is presented in detail only for $\alpha=2$, even valued $m$, and $\frac{m}{2} < C < \frac{3m}{2}$. Details of the achievable scheme for the other cases will be provided in~\cite{partha3}. 

Now, when $C=0$, links corresponding to levels $[1:m]$ cannot be used for transmitting a user's own data, as they are not present at its receiver. Only the links corresponding to levels $[m+1:n]$ can be used for transmission.  Bits transmitted on these upper levels will be received cleanly at the unintended receiver. Hence, when $C=0$, it may not be possible to achieve a nonzero secrecy rate. However, when $C>0$, it is possible to achieve a nonzero secrecy rate, as explained below for the case where $m$ is even.

When $\frac{m}{2} < C < \frac{3m}{2}$, the achievable scheme uses transmission of random bits, interference cancelation and time sharing. The transmitters share a combination of random bits and data bits through the cooperative links. More specifically, 
both the transmitters share $\frac{m}{2}$ random bits along with $C_1 = C - \frac{m}{2}$ data bits. In the first time slot, transmitter $1$ sends the $m$ random bits ($d_i$ and $e_i$) on alternate levels in $[1:m]$. In order to eliminate the interference caused by these random bits at receiver $2$, the data bits of transmitter $2$ are precoded (xored) with these $m$ random bits and transmitted on the levels from $[m+1:2m]$ from transmitter $2$. The random bits are not canceled at receiver $1$. Further, receiver $1$ has no knowledge of these random bits. Hence, it cannot decode the bits intended to receiver $2$. Also, the data bits of transmitter $2$ received through the cooperative link are transmitted at the upper levels $[n-C_1+1:n]$ from transmitter $1$. Again, in order to ensure secrecy at receiver $1$, transmitter $2$ sends the same data bits at levels $[m-C_1+1:m]$ along with the $C_1$ data bits of transmitter $1$, also received through cooperation. This not only cancels the interference due to the bits sent on levels $[n-C_1+1:n]$ at receiver $1$, but also enables transmitter $2$ to relay the data bits of transmitter $1$. 

\begin{figure}
\centering
\mbox{\subfigure[][First slot:  $(R_1,R_2)=(2,3)$.] {\includegraphics[width=1.65in]{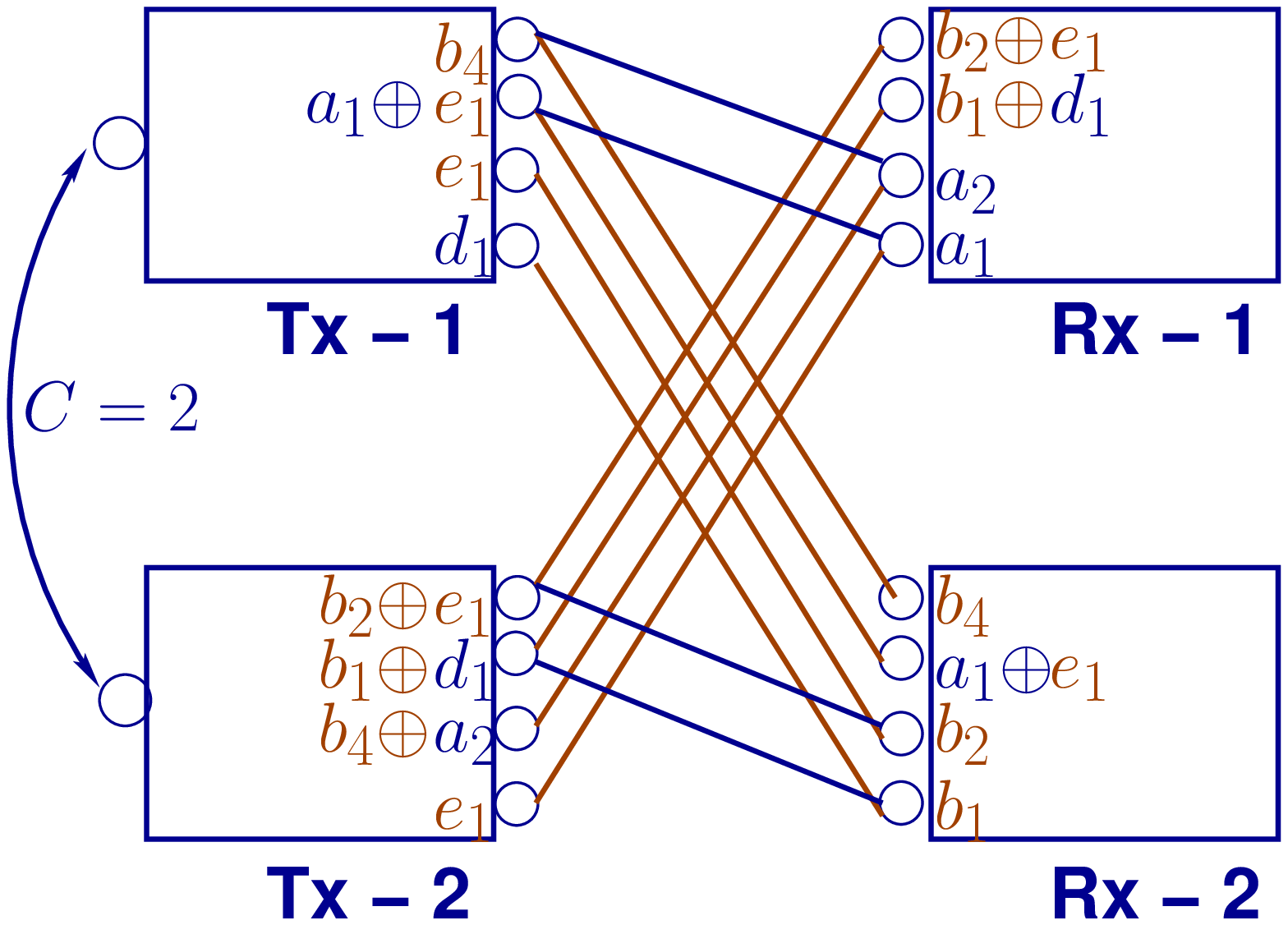}\label{fig:veryhigh7}} \quad
\subfigure[][Second slot: $(R_1,R_2)=(3,2)$.] {\includegraphics[width=1.65in]{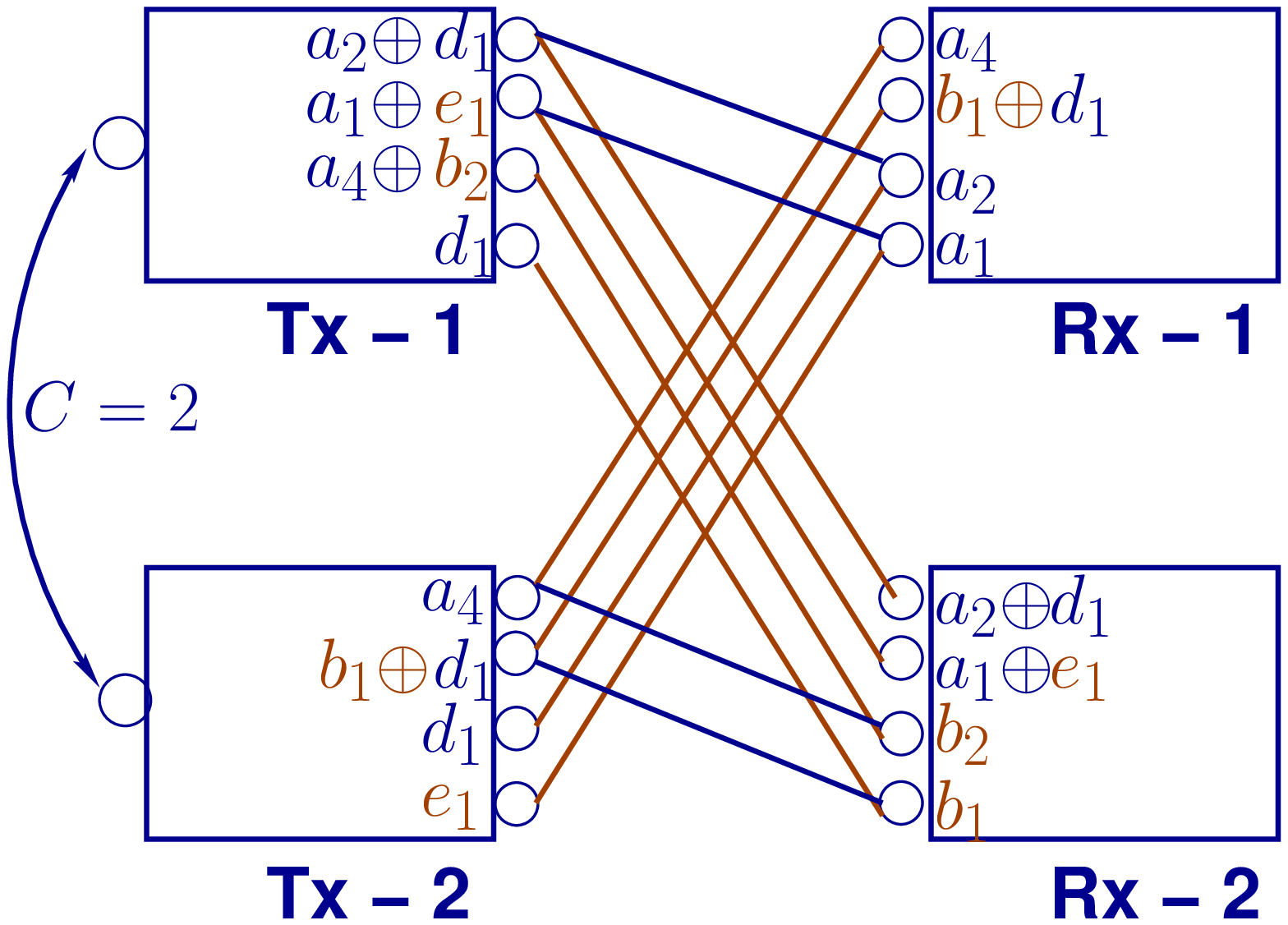}\label{fig:veryhigh8}}} \\
\caption[]{LDIC with $m=2$ and $n=4$: $C=2$ and $R_S=2.5$.}\label{fig:veryhigh9}
\pullUp
\end{figure}

In the remaining upper levels $[m+1:n-C_1]$, transmitter $1$  sends its own data bits xored with random bits. Transmitter $2$ transmits the same random bits on levels $[1:C_1]$ to cancel the random bits at receiver $1$. In this way, transmitter $1$ sends $m-C_1$ data bits of its own and $C_1$ data bits of transmitter $2$, in the first time slot. Simultaneously, transmitter $2$ is able to send $m$ data bits of its own and $C_1$ data bits of transmitter $1$. In the second time slot, the roles of transmitters $1$ and $2$ are reversed. Figures~\ref{fig:veryhigh7} and \ref{fig:veryhigh8} illustrate the scheme for $m=2$ and $n=4$, with $C=2$ bits. It can be observed that, in this case, it is possible to securely transmit $2.5$ bits per user. 

When $0 < C \leq \frac{m}{2}$, interestingly, transmitters share only random bits through the cooperative links. The achievable scheme involves transmitting the data bits xored with the random bits. The same random bits are transmitted by the other transmitter, so as to cancel them out at the desired receiver. When $\frac{3m}{2}  \leq C \leq n$, the achievable scheme uses interference cancelation, and transmitters share  only data bits through the cooperative links. The details are omitted due to lack of space.

\section{Discussion}\label{sec:discussion}
\subsection{Numerical Examples}\label{sec:numericexamples}
Now, some examples presented to illustrate the optimality of the proposed achievable scheme for different $m,n$ and $C$. The achievable secrecy rate per user is plotted against the capacity of the cooperative link for different values of $m$ and $n$. Also plotted is the per user capacity of the SLDIC with transmitter cooperation, but without the secrecy constraints\cite{wang1}. 

\begin{figure}[t]
\begin{center}
\setxysizeo
\epsffile{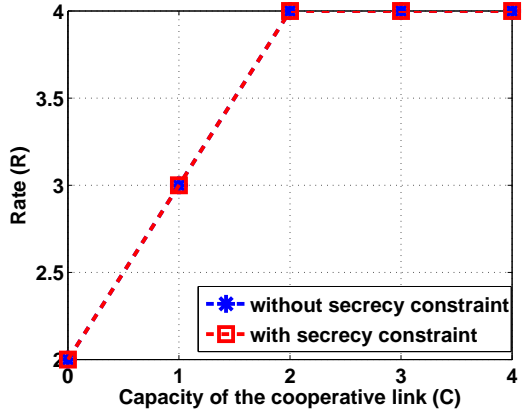}
\caption{Achievable rate of the LDIC with $m=4$ and $n=2$.} 
\label{fig:compare1}
\end{center}
\pullUp
\end{figure}
\begin{figure}[t]
\begin{center}
\setxysizeo
\epsffile{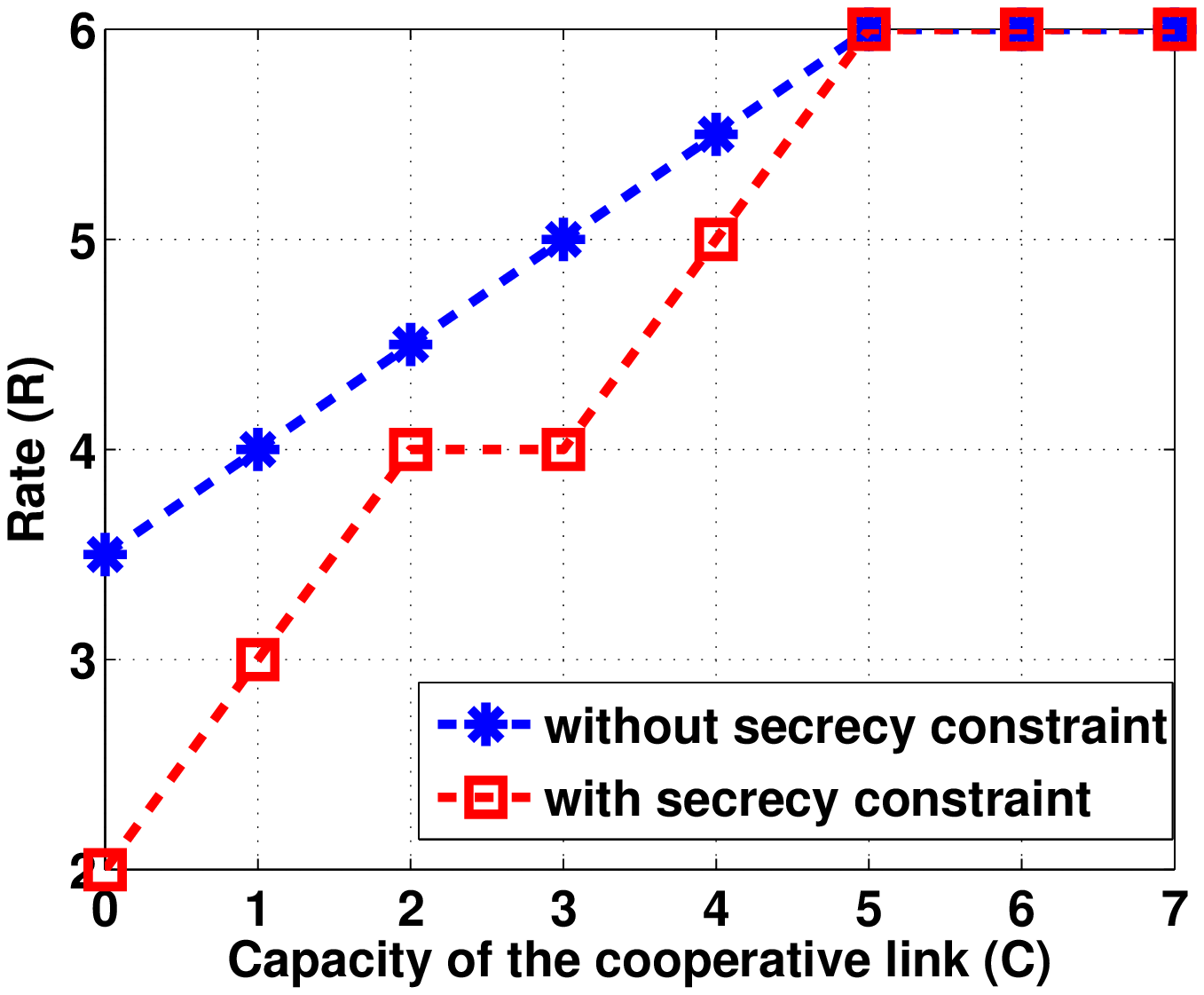}
\caption{Achievable rate of the LDIC with $m=6$ and $n=5$.} 
\label{fig:compare2}
\end{center}
\pullUp
\end{figure}
\begin{figure}[t]
\begin{center}
\setxysizeo
\epsffile{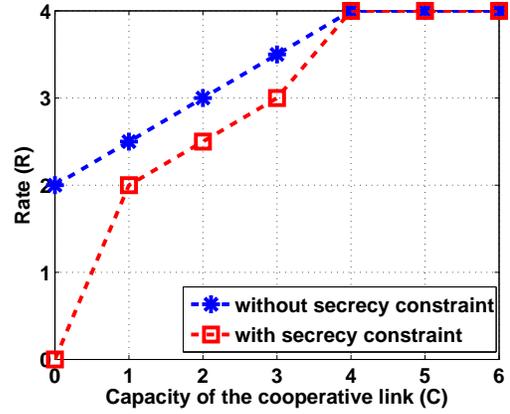}
\caption{Achievable rate of the LDIC with $m=2$ and $n=4$.} 
\label{fig:compare6}
\end{center}
\pullUp
\end{figure}

In Fig. \ref{fig:compare1}, the achievable secrecy rate in (\ref{weakach2}) is plotted for the $m=4$, $n=2$ case. It can be seen that the proposed scheme meets the capacity without the secrecy constraint for all values of $C$. Hence, the proposed scheme is optimal for $m=4$ and $n=2$, and, interestingly, one obtains secrecy for free. In this case, the achievable scheme is based on interference cancelation. When $m=6$ and $n=5$, the achievable rate in (\ref{eq:weakmod3}) is plotted in Fig. \ref{fig:compare2}. Here, the scheme is optimal for $C\geq 5$. There is a gap between the achievable rate and the capacity without the secrecy constraint when $C < 5$. This could, perhaps, be due to the secrecy constraint itself. The derivation of outer bounds on the achievable rate, that would settle this question, is work in progress.
In Fig.~\ref{fig:compare6}, the achievable secrecy rate is plotted as a function of $C$, with $m=2$ and $n=4$. It can be observed that the proposed scheme is optimal when $C \geq 4$. 

It is interesting to note that although the users are not allowed to decode the other user's message,  it is possible to achieve nonzero secrecy rate with cooperation, even in the very high interference regime~$(\alpha \geq 2)$. Thus, with cooperation ($C>0$), it is possible to achieve nonzero secrecy rate in almost all cases. Also, when $C=n$, the proposed scheme achieves the maximum possible rate, i.e., $\max\{m,n\}$.

\subsection{Further Remarks}
\begin{enumerate}
\item When $(0 < \alpha \leq \frac{1}{2})$, the proposed scheme is optimal for all values of $C$. In this case, interference cancelation suffices to achieve the optimal rate.
\item In the moderate interference regime $(\frac{2}{3} < \alpha < 1 )$, the proposed scheme achieves nonzero secrecy rate for all values of $C$. It is possible to transmit data bits securely in the higher levels by intelligently choosing the placement of data and random bits, in addition to interference cancelation.
\item Unlike in the weak interference regime, the achievable scheme in the moderate and high interference regimes uses random bit transmission, when~$C=0$.
\item In the very high interference regime $(\alpha \geq 2)$, it is not possible to ensure secrecy with random bit transmission without  cooperation. However, with cooperation, it is possible to achieve nonzero secrecy rate. Further, the scheme may use the sharing of random bits and/or data bits, depending on $m$, $n$, and $C$. 

\item In all the interference regimes, the presented scheme always achieves nonzero secrecy rate with cooperation, except for the $\alpha = 1$ case.
\item When $C=n$ and $\alpha \neq 1$, i.e., when the cooperative link is as strong as the strength of the interference, the proposed scheme achieves the maximum possible rate of~$\max\{m,n\}$. 
\end{enumerate}
\section{Conclusions}
This work proposed novel achievable schemes for the $2$-user symmetric deterministic interference channel with transmitter cooperation. The achievable scheme used a combination of interference cancelation, random bit transmission, relaying of the other user's data bits, and time sharing, depending on the values of $\alpha$ and $C$. Several interesting results were obtained from the proposed achievable schemes. For example, when $\frac{2}{3}<\alpha <1$ and $1 < \alpha < 2$, random bit transmission helps ensure secrecy. With further increase in the strength of the interference $(\alpha \geq 2)$, random bit transmission is rendered ineffective. But, with cooperation, it is possible to achieve a nonzero secrecy rate, even when the interference is very strong. Finally, when $0 < \alpha \leq \frac{1}{2}$, the achievable scheme is found to be optimal for all values of $C$. When $\alpha \geq 2$, sharing random bits, or data bits, or both, outperforms sharing only data bits through the cooperative links. Finding outer bounds and extending the results to the Gaussian setup are interesting avenues for future work.
\bibliographystyle{IEEEtran}
\bibliography{IEEEabrv,bibJournalList,refs}
\end{document}